\documentclass{article}
\usepackage{spconf,amsmath,graphicx}
\usepackage{subcaption}
\usepackage{hyperref}

\title{MAEVI: Motion Aware Event-Based Video Frame Interpolation}
\name{Ahmet Akman, Onur Selim Kılıç, A. Aydın Alatan\thanks{The authors would like to thank METU Center for Image Analysis (OGAM) for partially funding this research. The code and pre-trained model can be reached from the following link: \href{https://github.com/ahmetakman/MAEVI}{github.com/ahmetakman/MAEVI}}}
\address{Dept. of Electrical-Electronics Engineering, Cent. for Image Analysis (OGAM), METU, 06800 Ankara, TR}
\begin{document}
%
\maketitle
\begin{abstract}
Utilization of event-based cameras is expected to improve the visual quality of video frame interpolation solutions. 
We introduce a learning-based method to exploit moving region boundaries in a video sequence to increase the overall interpolation quality.
Event cameras allow us to determine moving areas precisely; and hence, better video frame interpolation quality can be achieved by emphasizing these regions using an appropriate loss function. 
The results show a notable average \textit{PSNR} improvement of $1.3$ dB for the tested data sets, as well as  subjectively more pleasing visual results with less ghosting and blurry artifacts.  
\end{abstract}
\section{Introduction}
\label{sec:intro}

Video frame interpolation (VFI) is a popular video processing problem in which a new frame is generated to fill the temporal gap between existing frames for the videos which have a lower frame-rate. The complexity of motions of the objects in the scene and the large displacements of the camera motion still make VFI problem a challenging issue to solve.

Majority of the past VFI methods use only the visual information from the original video frames to estimate the motion of scene points mostly via optical flow estimation. However, this approach can be problematic at relatively low frame-rates or high motion, since the inter-frame motion might not be estimated precisely, when there is significant displacement or non-linear object trajectories. 
While high frame-rate visual sensors might provide a solution, such cameras are often quite expensive.

Event cameras offer a unique solution to the problems encountered by VFI. Unlike the conventional cameras that capture frames from absolute scene brightness, the  event cameras only measure changes in the brightness at each pixel to produce a stream of events \cite{Survey}. This stream includes information only about the position, time, and polarity changes, while providing a high temporal resolution on the order of microseconds. These properties allow precise capture of the trajectories of fast-moving objects via event cameras.

A recent method, namely EVFIA \cite{evfia}, exploits event-based data for high quality VFI by fusing the image of a conventional camera with an event-based stream. 
In \cite{evfia}, the fusion of these two sensors is obtained by deformable convolutions \cite{Deformable}, which are end-to-end mechanisms capable of learning various geometric transformations. 

\textbf{Motivation.} Although EVFIA \cite{evfia} outperforms both pure image-based algorithms as well as event-based methods for FVI, this algorithm still has several shortcomings. 
First of all, the interpolated images contain some ghosting and/or blur artifacts due to suboptimal trajectory estimates, especially around motion boundaries. 
Moreover, the color content of some objects in the interpolated frames is not accurate, since these pixels have a tendency to replicate the color of their surrounding objects, creating hollow silhouettes, especially for the cases where the background has different colors.

For any learning-based VFI method, such as EVFIA \cite{evfia}, the most important drawback for the interpolation architecture is the lack of information about the precise boundary between the independently moving objects and the background scene pixels. 
However, the event cameras bring significant advantages to sense and capture the positions of moving and stationary pixels, compared to the ill-posed nature of moving object segmentation problem for the conventional cameras. 
This extra event information makes both training and inference steps to perform better.
By only considering the conventional video, it is challenging to precisely segment fast-moving or merely-seen objects in the scene. 
The event-based camera provides an alternative for implicitly determining and exploiting such regions, since events are usually located around the contours of the moving objects. 

Our main contributions can be summarized as follows:
\vspace{-8pt}
\begin{itemize}
    \item We introduce MAEVI, a critical extension to a recent event-based VFI algorithm \cite{evfia} that incorporates spatial motion awareness.\vspace{-8pt}
    \item We propose a novel moving region filter to weight the input frames depending on their motion content by using event information.\vspace{-8pt}
    \item We suggest a new loss function for VFI problem to ensure that the pixels with motion have significant influence during the training process.
    \vspace{-8pt}
\end{itemize}

\begin{figure*}[htbp!] 
    \centering{\includegraphics[width=0.8\linewidth]{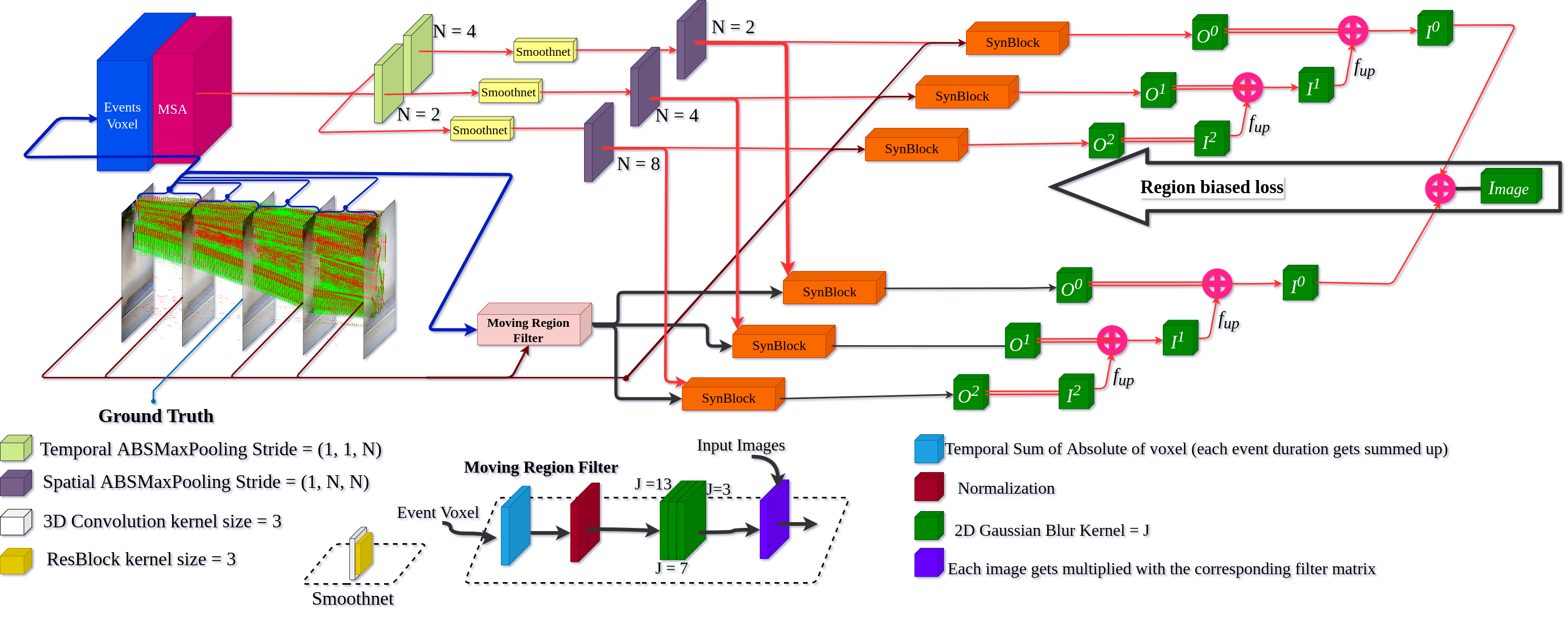}}
    \vspace{-10pt}
    \caption{Block diagram of the proposed architecture.} 
    \label{archFigure}
\end{figure*}

\section{Related Work}
\label{sec:relatedwork}
VFI problem has been extensively studied with many established solutions. Recent learning-based methods for VFI can be categorized into three groups: flow-based techniques \cite{DAIN, BMBC, qvi_nips19,sim2021xvfi, Lu_2022_CVPR,huang2022rife}, kernel-based approaches \cite{Shi_2022_CVPR,Lee2020AdaCoFAC, Niklaus_CVPR_2017, Niklaus2021} and phase-based methods \cite{meyer2018phasenet}.

Utilization of event cameras can be considered as a new direction for improving VFI visual quality, as they offer a higher dynamic range capability, low data-rate, and low-latency temporal information. Several studies \cite{evfia,Tulyakov_2022_CVPR,Tulyakov_2021_CVPR,WeaklySupervised,EvIntSR,Wang_2021_ICCV} have attempted to address VFI issues by utilizing information from event-based cameras. 

In \cite{Tulyakov_2021_CVPR}, motion flow masks for intermediate frames are estimated from event voxels, then, some artificial frames are generated by warping adjacent images by the estimated flow, and an attention averaging module is used to fuse three sister images obtained from the warping refinement module. 

In a two-stage method \cite{WeaklySupervised}, the events are fused with images in the first stage. The fused vectors are evaluated and integrated by a subpixel transformer network in the next stage. 

The results in \cite{Tulyakov_2021_CVPR} are further improved by introducing a motion spline estimator and multi-scale feature fusion modules as in \cite{Tulyakov_2022_CVPR}. The motion spline estimator enables the pipeline to consider the continuous nature of movements in the scene. Unfortunately, this approach makes event representation more blurry due to the limited memory of the network. 

The aforementioned approach \cite{evfia} uses the attention mechanism to evaluate event information and fuse it with the frames by using deformable convolution blocks. Although this method \cite{evfia} significantly improves the blur by providing crispier images, it still suffers from ghosting effects. 


\section{Proposed Method}
\label{sec:method}
Let the frame to be interpolated be denoted as \(I_0\), while the original input frames are designated as  \(I_{-2}\), \(I_{-1}\), \(I_{1}\), and \(I_{2}\). The event intervals between these frames are indicated by \(E_{-2 \to -1}\), \(E_{-1 \to 0}\), \(E_{0 \to 1}\), and \(E_{1 \to 2}\), which are represented by using voxels whose dimensions are $4 \times N_{TB} \times H \times W$ ($N_{TB}$ indicates the number of time bins within an event interval, and $H$ and $W$ represent height and width, respectively).

 The proposed framework in Figure \ref{archFigure} consists of two stages. The first stage generates the feature maps closely following the architecture presented in E-VFIA \cite{evfia}. At this stage, additional to \cite{evfia}, a novel \textit{moving region filter} paradigm, which tries to localize moving pixels from event voxels, is utilized. The second stage is built to generate the intermediate frame by utilizing input frames and the feature maps created in the first stage. In this second stage, there are two sister structures that utilize three synchronization blocks, i.e. \textit{SynBlocks} \cite{Shi_2022_CVPR} whose inputs have different dimensions. One of these structures is directly fusing standard images with events, whereas the other sister explicitly considers the movement in the scene by the moving region filters. Finally, using both of the outputs from these  structures, a motion-aware loss function is adopted to obtain the final output.

\textbf{Input stage and feature map generation.}
At the input stage, the events are converted into three-dimensional volumes through the voxel-grid representation   \cite{Tulyakov_2021_CVPR} whose dimensions are $4 \times N_{TB} \times H \times W$. Similar to \cite{evfia}, a multi-head self-attention mechanism is initially employed. Then, three feature maps are generated through consecutive operation of temporal absPooling, SmoothNet, and spatial absPooling. AbsPooling layers are max-pooling layers \cite{evfia} that consider the absolute of the original entries for max-pooling operation. Furthermore, SmoothNet blocks are composed of a simple 3D convolution followed by a Res-Net block, as used in \cite{Shi_2022_CVPR} to smooth out the vector determined by noisy event data. 

\textbf{Moving region filter.}
For VFI, the most informative part of the image frames consists of the pixels that are moving in the scene. The motion region filter is introduced to emphasize the parts of the image changing due to motion. This filter utilizes the voxel grids in a simple manner. First, an absolute sum is performed on the temporal dimensions of event voxels to prevent positive and negative events from canceling each other. In other words, the event volumes (a.k.a. event-voxels)  $E_{-2}, E_{-1}, E_{1}$, and $E_{2}$ are reduced to a volume with dimensions $4 \times H \times W$. Secondly, a standard normalization operation is performed to avoid having large numbers in the spatial dimension. 

In general, the events obtained by an event camera indicate contours of the moving objects. These events may not be sufficient to represent the moving object in the scene, and some of the pixels belonging to the moving object are still inactive. In order consider the amount of information in the aforementioned inactive pixels, a series of 2D Gaussian filters 
are used on the normalized vectors. The resulting vectors act as filters and multiplied with the input image before being fed to SynBlock for VFI.

\textbf{Fusion with standard frames.}
Deformable convolutions provide a straightforward and effective mechanism for enabling CNNs to learn a range of geometric transformations based on the input data. 
Deformable convolutions are proposed in \cite{Lee2020AdaCoFAC} that are useful in VFI for the scenarios in which the scene includes large relative motions. 
The algorithm presented in \cite{Shi_2022_CVPR}, which only utilizes frames to interpolate intermediate images, extends this concept by exploiting the spatial-temporal domain via vision transformers. 
In \cite{evfia}, utilization of the deformable convolutions is demonstrated  to fuse event-originated feature maps with the standard images. 
We also adapted SynBlocks from \cite{evfia}, which has deformable convolutions as their essence. 
The fusion of standard images with event information is performed using three similar SynBlocks, and they generate different scaled interpolated images. 
The resolutions of the small images are increased, and the output images are summed to create an output.

\textbf{Fusion with region filtered frames.}
Parallel to the fusion with standard frames, three other SynBlocks are used that input the moving region filter outputs. These SynBlocks synthesize interpolated frames from stationary background suppressed frames.


\textbf{Loss function.}
Past VFI research reveals that the artifacts in the interpolated images are generally observed within the areas with large motions. 
Since the events indicate the regions of the scene with motion, the loss function should take these regions into consideration. 
Therefore, another region filter, namely the loss filter, is constructed by averaging filters of the nearest images. And this filter is used in the loss function.

The total loss function consists of two different loss calculations. First, a regular $L1$ loss from ground truth and the interpolated image is obtained and denoted as $L_{full}$. Second, another $L1$ loss is calculated using ground truth and interpolated images after they are filtered with the aforementioned loss filter. This second loss is denoted as $L_{filtered}$. and the overall loss is calculated as $Loss = \alpha L_{filtered} + (1-\alpha) L_{full}$
Here, the value $\alpha$ is selected as 0.6 via experimentation. 
In other scenarios, the loss values decrease rapidly due to the stationary background in the scene. 
This definition of loss gives more focus to hard-to-learn areas and prevents rapid loss drop during training. 
In other words, the training process is biased towards the moving domains of the scene.

\begin{figure*}[!t]
	\begin{center}
	\begin{minipage}[h]{0.120\linewidth}
		\centering
		\includegraphics[width=\linewidth]{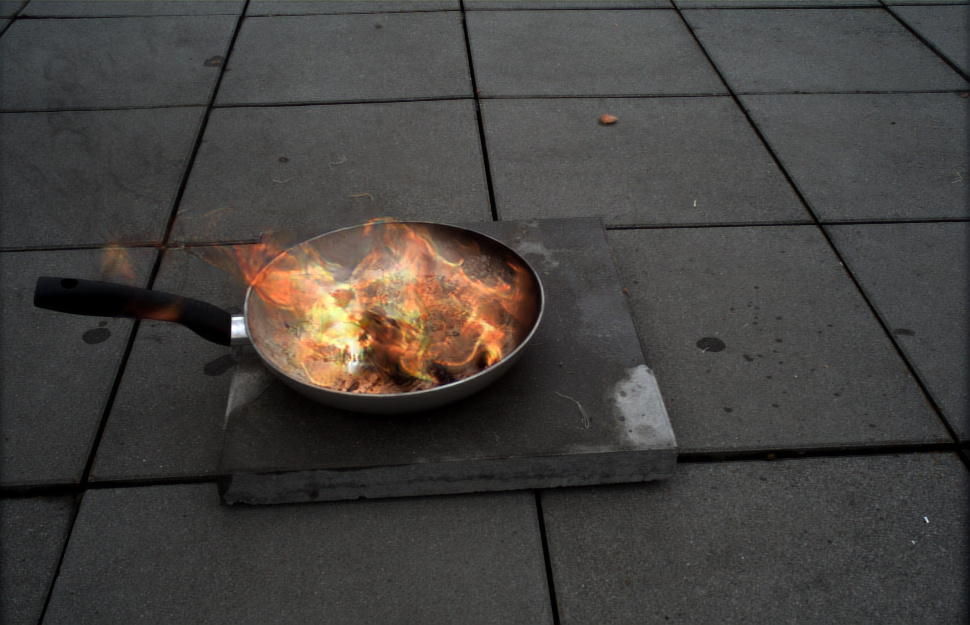}
	\end{minipage}
	\begin{minipage}[h]{0.120\linewidth}
		\centering
		\includegraphics[width=\linewidth]{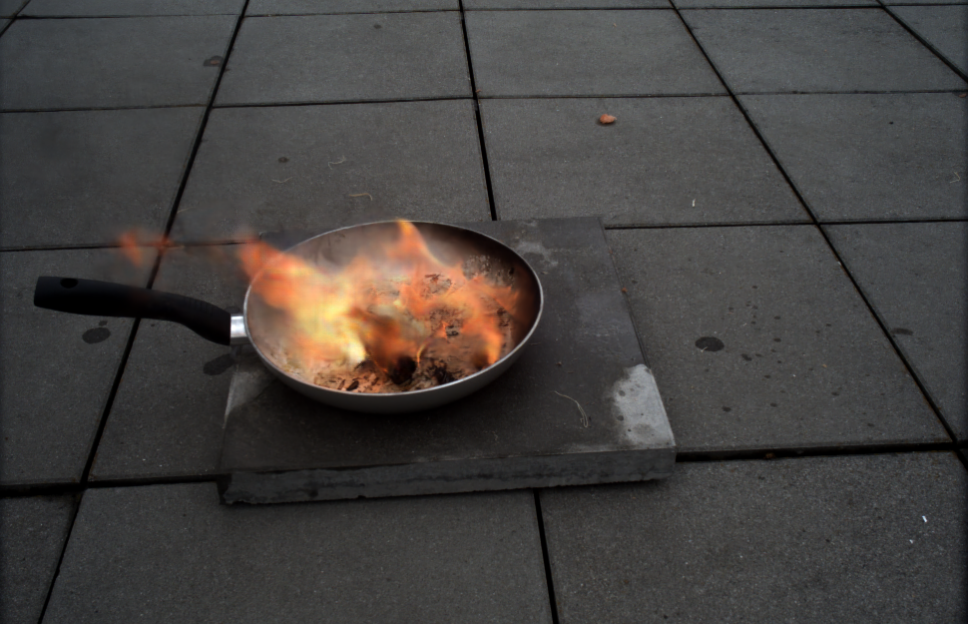}
	\end{minipage}
	\begin{minipage}[h]{0.120\linewidth}
		\centering
		\includegraphics[width=\linewidth]{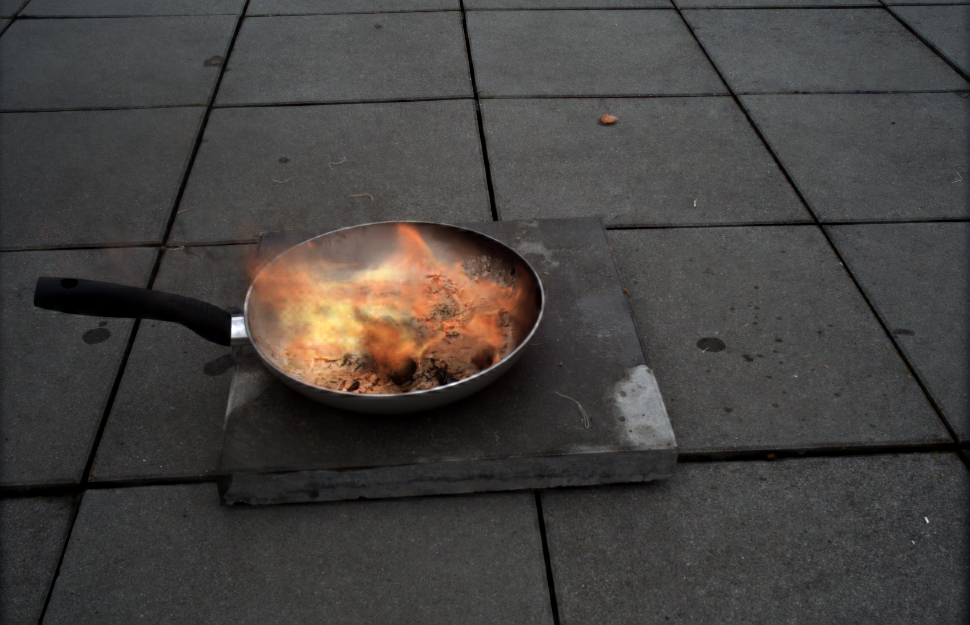}
	\end{minipage}
	\begin{minipage}[h]{0.120\linewidth}
		\centering
		\includegraphics[width=\linewidth]{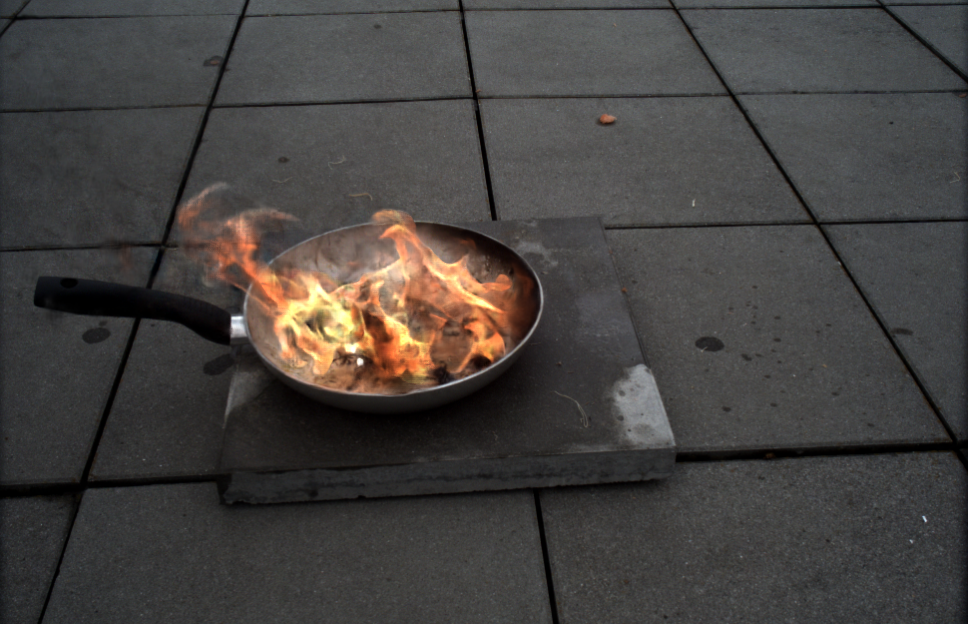}
	\end{minipage}
 	\begin{minipage}[h]{0.120\linewidth}
		\centering
		\includegraphics[width=\linewidth]{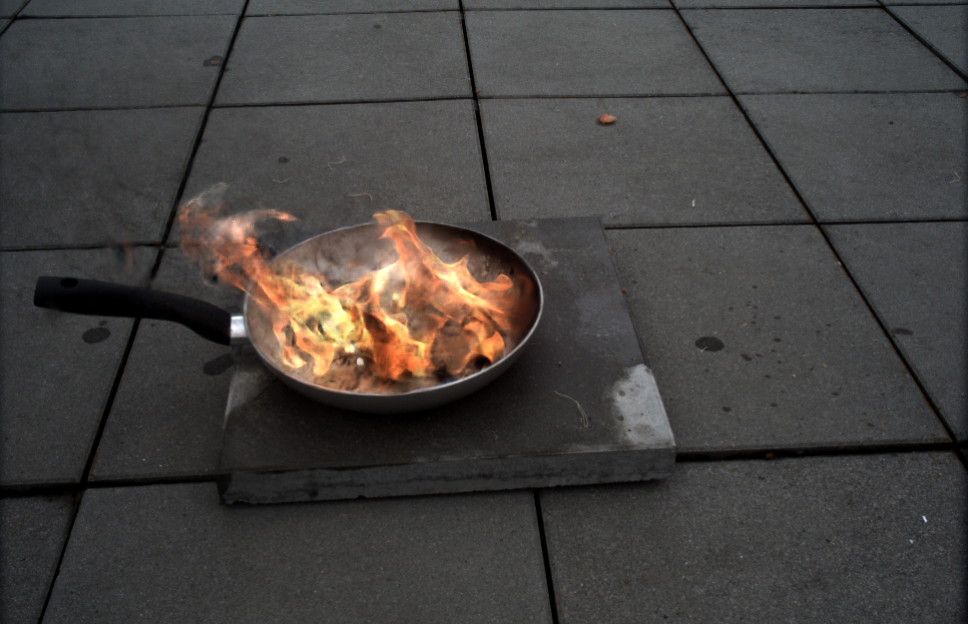}
	\end{minipage}
	\begin{minipage}[h]{0.120\linewidth}
		\centering
		\includegraphics[width=\linewidth]{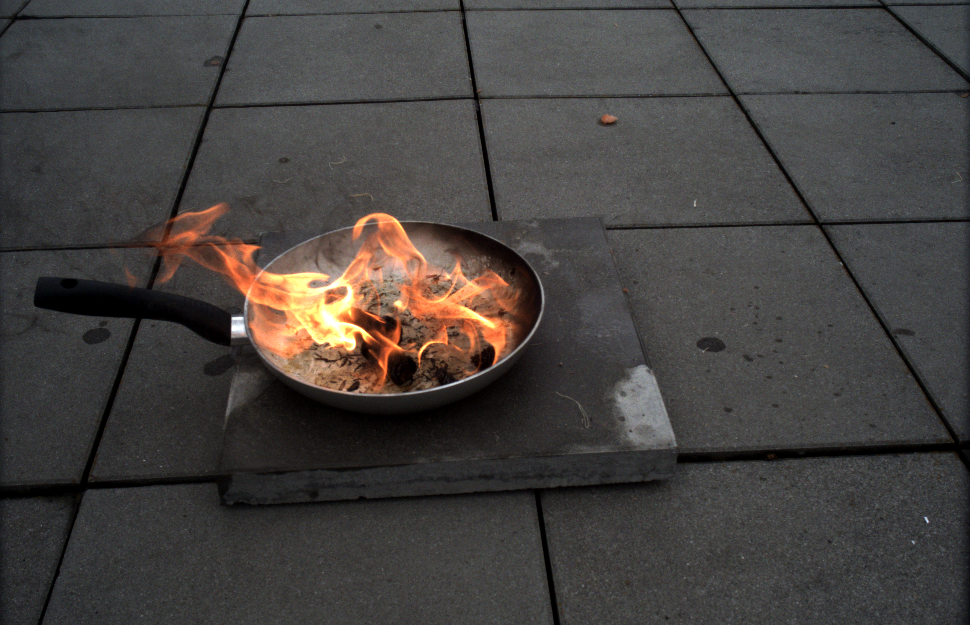}
	\end{minipage}
    \end{center}

    \begin{center}
	\begin{minipage}[h]{0.120\linewidth}
		\centering
		\includegraphics[width=\linewidth]{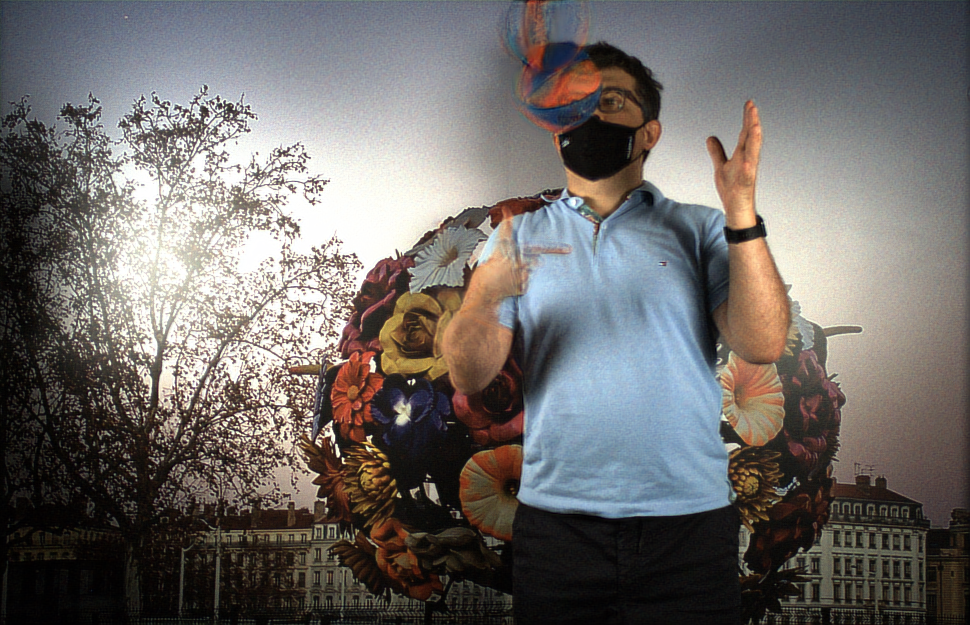}
		\scriptsize{Time Lens \cite{Tulyakov_2021_CVPR}}
	\end{minipage}
	\begin{minipage}[h]{0.120\linewidth}
		\centering
		\includegraphics[width=\linewidth]{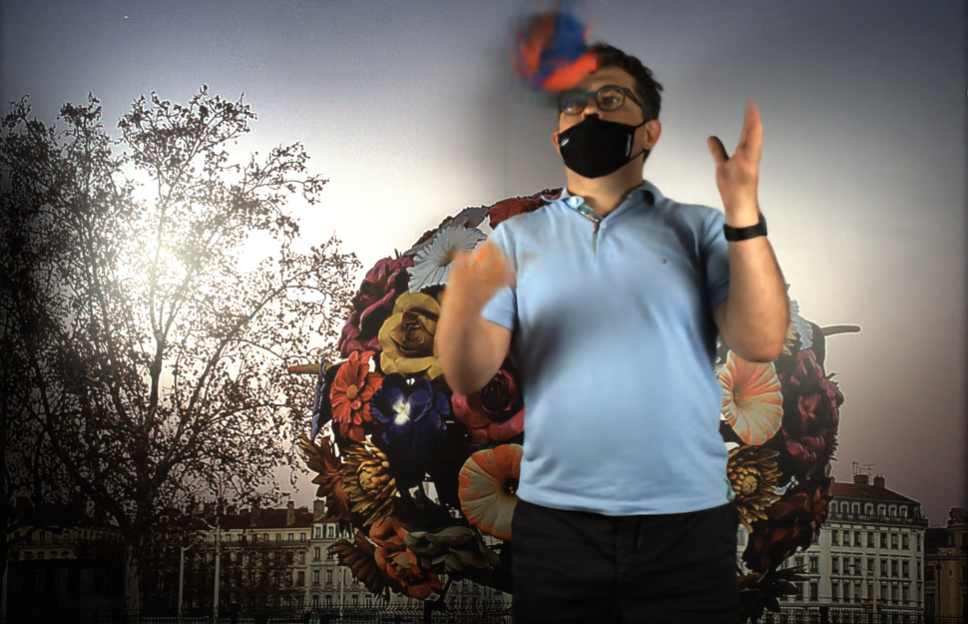}
		\scriptsize{FLAVR \cite{Kalluri2020FLAVRFV}}
	\end{minipage}
	\begin{minipage}[h]{0.120\linewidth}
		\centering
		\includegraphics[width=\linewidth]{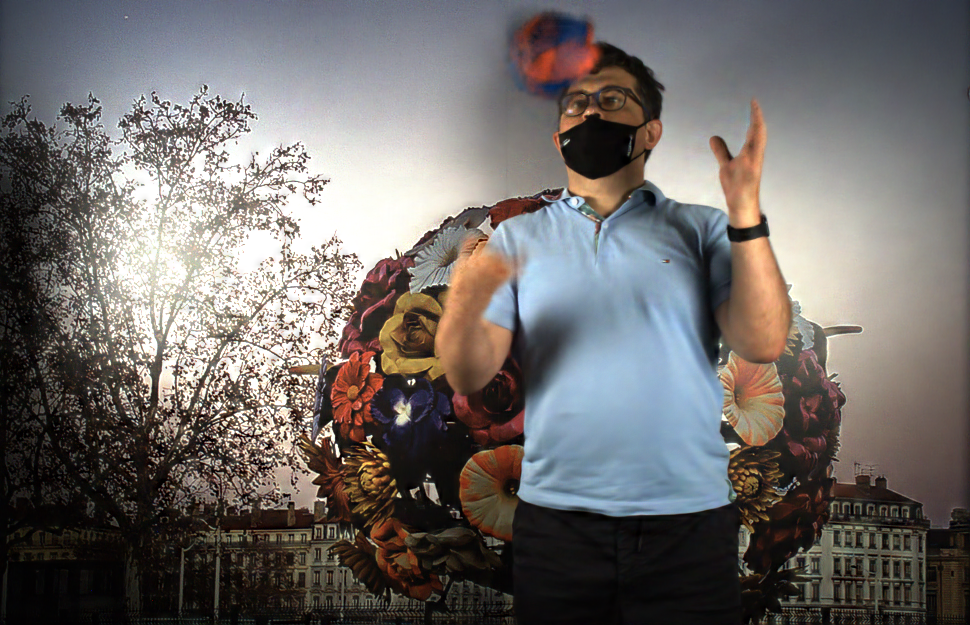}
		\scriptsize{VFIT \cite{Shi_2022_CVPR}}
	\end{minipage}
	\begin{minipage}[h]{0.120\linewidth}
		\centering
		\includegraphics[width=\linewidth]{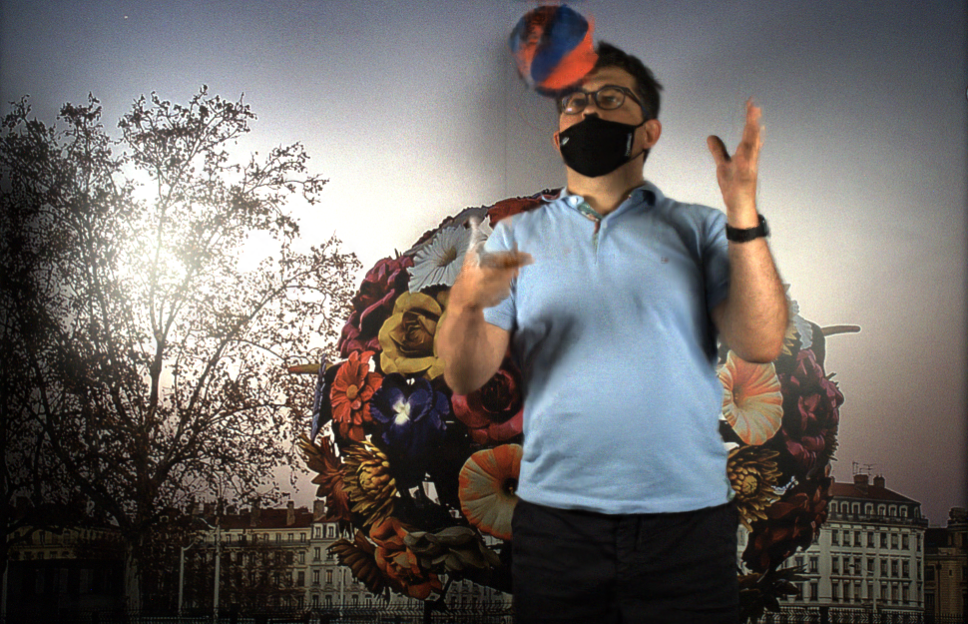}
		\scriptsize{E-VFIA\cite{evfia}}
	\end{minipage}
 	\begin{minipage}[h]{0.120\linewidth}
		\centering
		\includegraphics[width=\linewidth]{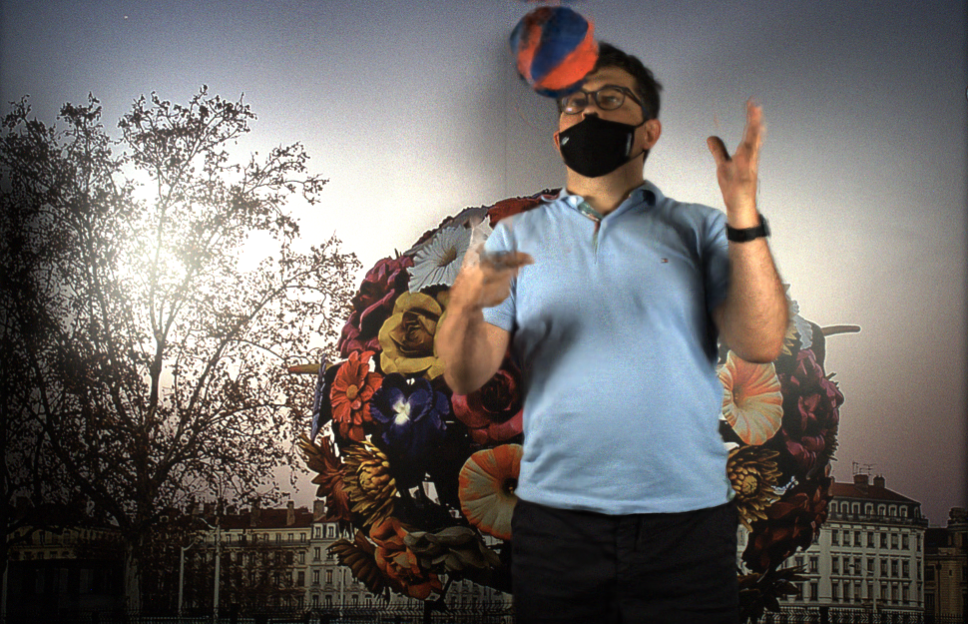}
		\scriptsize{\textbf{Ours}}
	\end{minipage}
	\begin{minipage}[h]{0.120\linewidth}
		\centering
		\includegraphics[width=\linewidth]{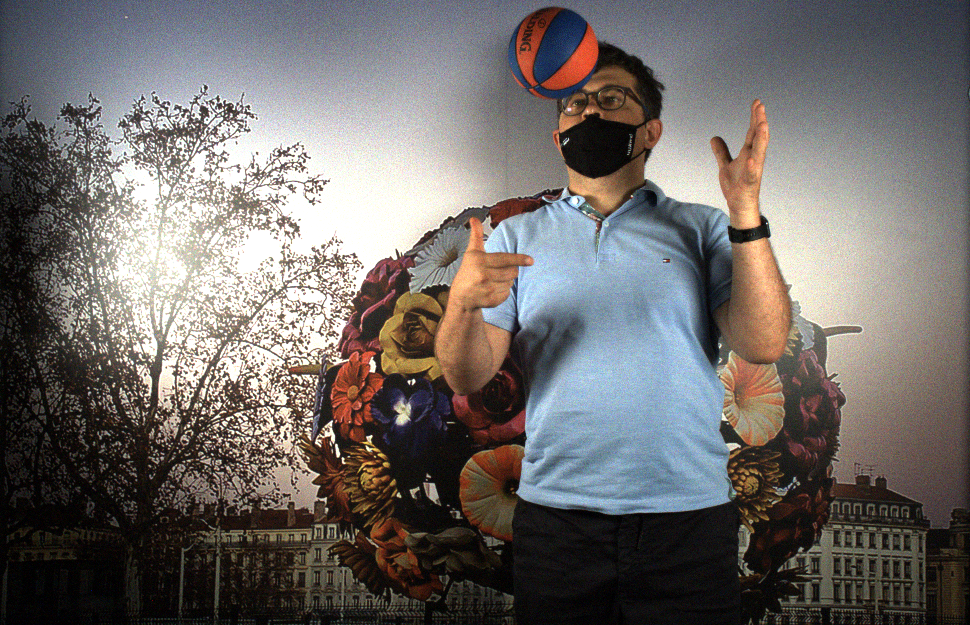}
		\scriptsize{GT}
	\end{minipage}
    \end{center}
\vspace{-4mm}
	\caption{Qualitative evaluations between the state-of-the-art methods and the proposed algorithm}
	\label{fig:qualitative_comparisons}
\end{figure*}

\begin{figure*}
\begin{subfigure}{.195\textwidth}
      \centering
      \includegraphics[width=.9\linewidth]{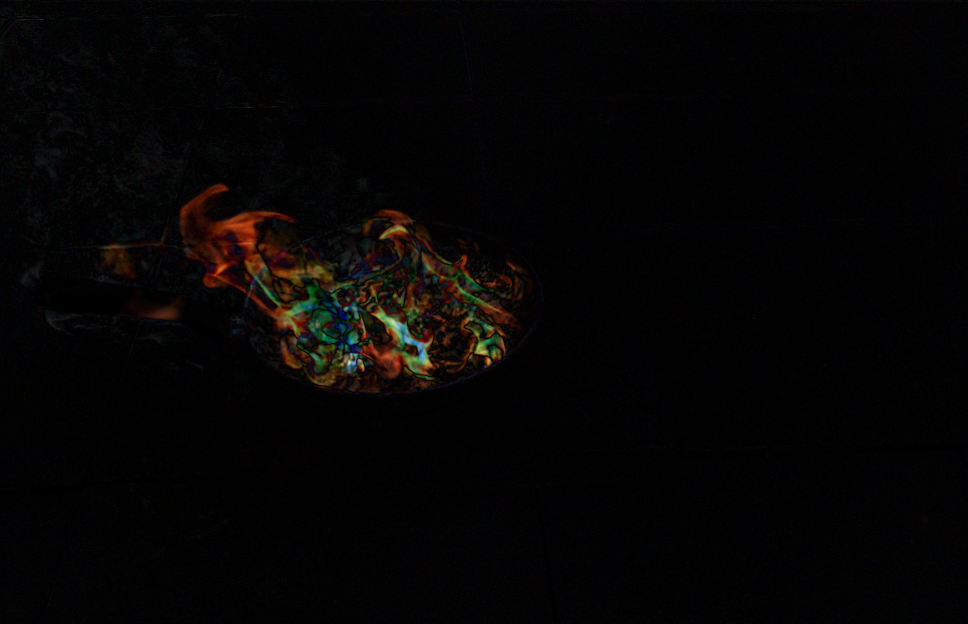}
      \caption{Time Lens \cite{Tulyakov_2021_CVPR}}
      \label{fig:sfig2}
\end{subfigure}
\begin{subfigure}{.195\textwidth}
    \centering
    \includegraphics[width=.9\linewidth]{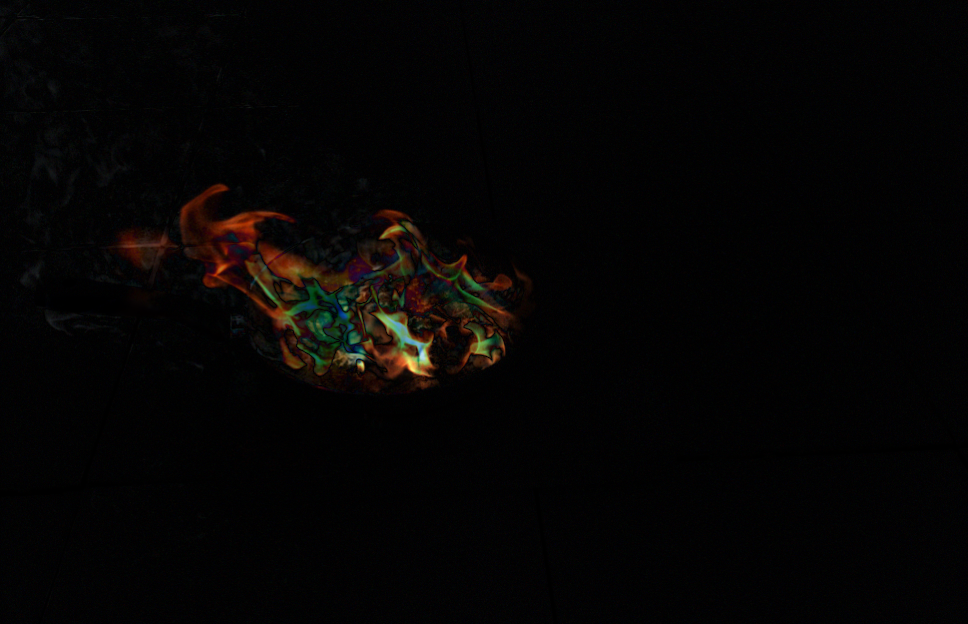}
    \caption{FLAVR \cite{Kalluri2020FLAVRFV}}
    \label{fig:sfig4}
\end{subfigure}
\begin{subfigure}{.195\textwidth}
    \centering
    \includegraphics[width=.9\linewidth]{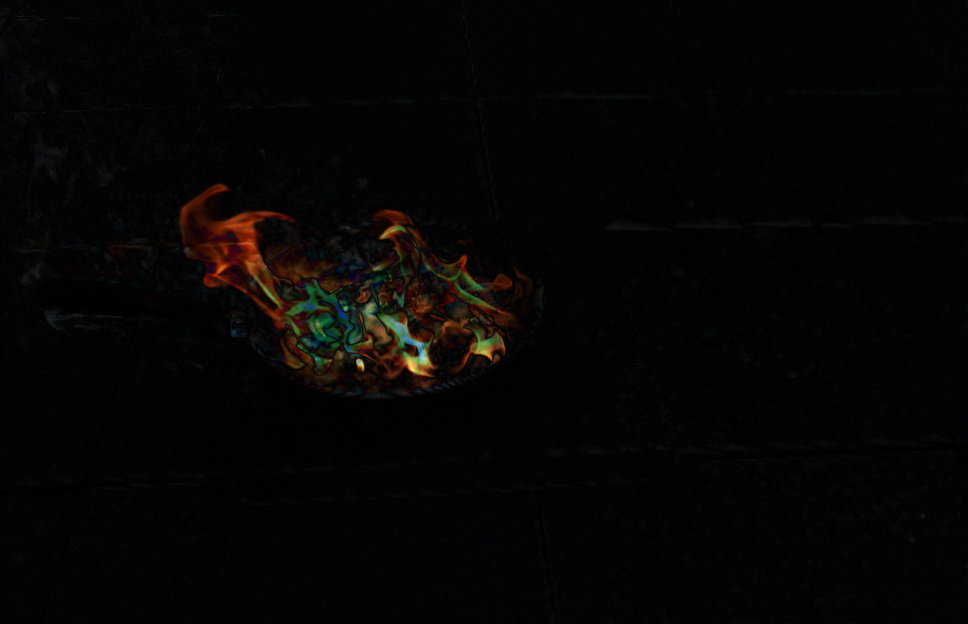}
    \caption{VFIT \cite{Shi_2022_CVPR}}
    \label{fig:sfig5}
\end{subfigure}
\begin{subfigure}{.195\textwidth}
    \centering
    \includegraphics[width=.9\linewidth]{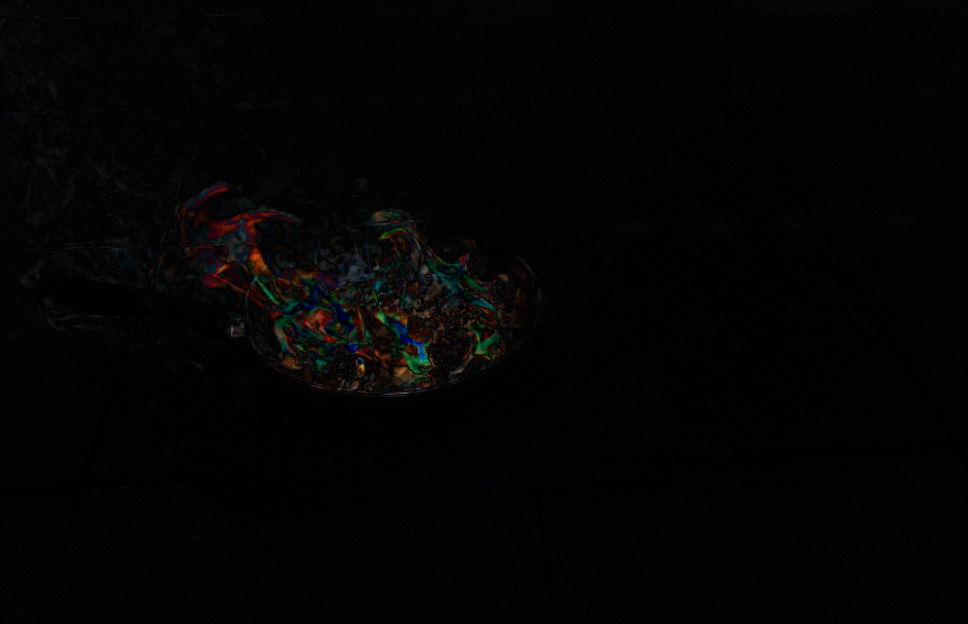}
    \caption{E-VFIA\cite{evfia}}
    \label{fig:sfig6}
\end{subfigure}
\begin{subfigure}{.195\textwidth}
    \centering
    \includegraphics[width=.9\linewidth]{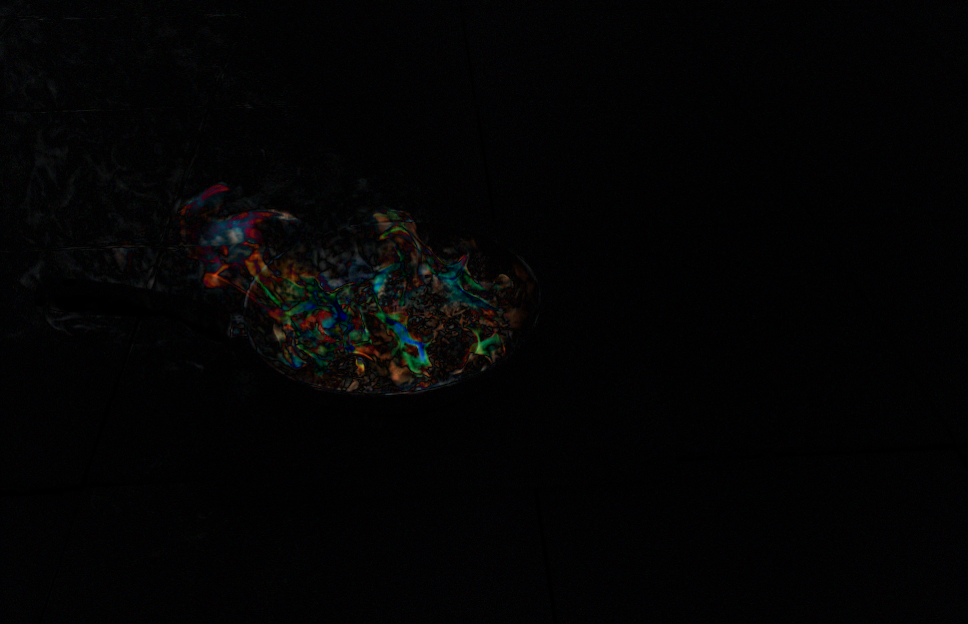}
    \caption{Ours}
    \label{fig:sfig7}
\end{subfigure}
\caption{Pixel-wise differences on the fire sample.}
\label{fig:fig}
\end{figure*}

\section{Experiments and Results}
\label{sec:results}
\subsection{Implementation Details} The training and test procedures are conveyed using PyTorch software framework. The learning rate is reduced gradually, starting from 0.0016. We employed the AdaMax optimizer, as in \cite{AdaMax}, with $\beta_1$ and $\beta_2$ values of 0.9 and 0.999, respectively. Our training batch size was 4, and we conducted the training on a workstation equipped with two 2080TI GPUs. The training is completed in 60 epochs. In order to accommodate memory constraints, we used images and associated events that were resized to $256 \times 256$ pixels. To ensure accurate quantitative results, we compared our proposed method using both full-scaled and downscaled frames with a resolution of $256 \times 256$ pixels. The selection of image resolution might significantly impact the obtained results.

\begin{table}[htbp]
    \caption{Comparisons for 256x256 setup in BS-ERGB.} \vspace{-20pt}
    \begin{center}
        \begin{tabular}{|c|c|c|c|c|}
        \hline
        \textbf{Method} & \textbf{Input} &\textbf{\shortstack{ \#Param\\(M)}}& \textbf{PSNR (dB)}& \textbf{SSIM} \\
        \hline
        \shortstack{FLAVR \cite{Kalluri2020FLAVRFV}}&Frames& 42.4 &31.72 & 0.9469 \\
        \hline
        VFIT \cite{Shi_2022_CVPR} &Frames& 29.0  & 32.08&  0.9449\\
        \hline
        \shortstack{Timelens \cite{Tulyakov_2021_CVPR} } &\shortstack{Frames\\Events}& 72.2 & 28.36 & 0.9320 \\
        \hline
        \shortstack{Timelens++ \\ \cite{Tulyakov_2022_CVPR}} &\shortstack{Frames\\Events}& 53.9 & 28.56 &  - \\
        \hline
        E-VFIA \cite{evfia} &  \shortstack{Frames\\Events} & \textbf{2.07} & 32.23 & 0.9581 \\
        \hline
        \textbf{Proposed}  &\textbf{\shortstack{Frames\\Events}}&3.02 & \textbf{33.57} &  \textbf{0.9618} \\
        \hline
        \end{tabular}
    \label{tabComp256}
\end{center}
\end{table}

\begin{table}[htbp]
    \caption{Comparisons for full-scale setup in BS-ERGB.} \vspace{-20pt}
    \begin{center}
        \begin{tabular}{|c|c|c|c|c|}
        \hline
        \textbf{Method} & \textbf{Input} &\textbf{\shortstack{ \#Param\\(M)}}& \textbf{PSNR (dB)}& \textbf{SSIM} \\
        \hline
        \shortstack{FLAVR \cite{Kalluri2020FLAVRFV}}&Frames& 42.4 &27.642 & 0.8729 \\
        \hline
        VFIT \cite{Shi_2022_CVPR} &Frames& 29.0  & 28.00 &  0.8767\\
        \hline
        \shortstack{Time lens \\ \cite{Tulyakov_2021_CVPR}} &\shortstack{Frames\\Events}& 72.2 & 23.97 & 0.7838 \\
        \hline
        \shortstack{Time lens++ \\ \cite{Tulyakov_2022_CVPR}} &\shortstack{Frames\\Events}& 53.9 & - &  - \\
        \hline
        E-VFIA \cite{evfia} & \shortstack{Frames\\Events} & \textbf{2.07} & 29.04 & 0.8771 \\
        \hline
        \textbf{Proposed} & \textbf{\shortstack{Frames\\Events}} & 3.025 & \textbf{29.27} & \textbf{0.8792} \\
        \hline
        \end{tabular}
    \label{tabCompFull}
\end{center}
\end{table}

\subsection{Quantitative Results} 
\begin{figure}[htbp]
    \vspace{-10pt}
    \centerline{\includegraphics[width=1.0\linewidth]{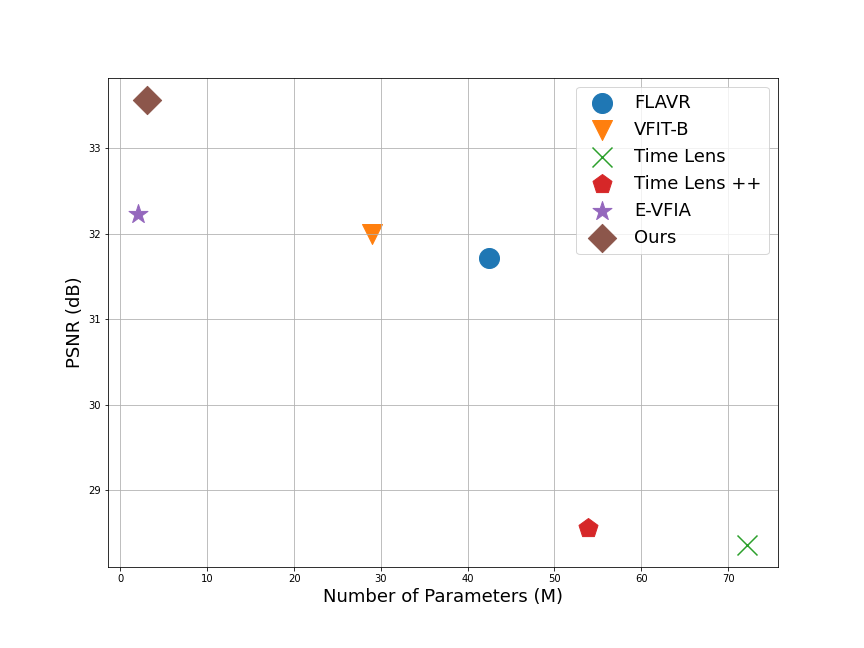}} \vspace{-15pt}
    \caption{Number of params. vs. PSNR values for different algorithms (FLAVR \cite{Kalluri2020FLAVRFV}, VFIT-B \cite{Shi_2022_CVPR}, Time Lens \cite{Tulyakov_2021_CVPR}, Time Lens++ \cite{Tulyakov_2022_CVPR}, E-VFIA \cite{evfia} and ours.):   for BS-ERGB \cite{Tulyakov_2022_CVPR}.} 
    \label{numParam}
\end{figure}

The proposed and compared methods are evaluated in BS-ERGB dataset \cite{Tulyakov_2022_CVPR}, where challenging scenarios with large movements form most of the dataset. Peak Signal to Noise Ratio (PSNR), and Structural Similarity Index (SSIM) metrics are  used during the quantitative evaluations. Moreover, the model sizes are presented in order to compare the mobility of the methods. Table \ref{tabComp256} and \ref{tabCompFull} exhibit the comparison of the proposed method versus state-of-the-art VFI methods in 256x256 and full-scale setups, respectively. Figure \ref{numParam} presents the graph between PSNR results and the model sizes in 256x256 setup. 
It can be observed that the proposed methods surpass state-of-the-art VFI methods quantitatively, improving PSNR 1.33 dB,  with a significantly small number of parameters.

\subsection{Qualitative Results} Qualitative comparisons are presented on BS-ERGB \cite{Tulyakov_2022_CVPR} dataset in full-scale resolution. Figure \ref{fig:qualitative_comparisons} show that the proposed method is less prone to ghosting effects and the output of the proposed method is more consistent than the other methods. Figure \ref{fig:fig} illustrates the pixel-wise consistency of the fire sample. It is observed that the proposed method enables smaller patches with a more homogeneous distribution of color channels.

\section{Conclusions} \label{sec:conclusion}
This paper introduces a novel approach to event-based video frame interpolation. Motion awareness is established through motion masks that are based upon event-based information. These masks are utilized by deformable convolutions during image generation to focus around moving areas. Moreover, a loss function based on motion awareness is defined to train the proposed motion-focused video frame interpolation network. The proposed method, MAEVI, improves the state-of-the-art, event-based, and frame-only methods in PSNR and SSIM metrics, still having a considerably small model size. 
Qualitatively, it is shown that the interpolated frames are less vulnerable to ghosting and blurry artifacts. 


\bibliographystyle{IEEEbib}
\bibliography{IEEEbib}

\end{document}